\documentclass[aps,12pt,onecolumn,titlepage,preprintnumbers,amsmath,amssymb]{revtex4}

\usepackage{graphicx}
\usepackage{dcolumn}
\usepackage{bm}

\begin{document}

\title{Probing energy barriers and quantum confined states of buried semiconductor heterostructures with ballistic carrier injection: An experimental study}

\author{Wei Yi}
\email{wyi@fas.harvard.edu}
\author{Venkatesh Narayanamurti}
\affiliation{Harvard School of Engineering and Applied Sciences, Harvard University, Cambridge, Massachusetts 02138
}

\author{Joshua M. O. Zide}
\author{Seth R. Bank}
\altaffiliation[Present address: Electrical and Computer Engineering Department, University of Texas, Austin, Texas 78758]{}
\author{Arthur C. Gossard}
\affiliation{Materials Department, University of California, Santa Barbara, California 93106
}


\begin{abstract}

A three-terminal spectroscopy that probes both subsurface energy barriers and interband optical transitions in a semiconductor heterostructure is demonstrated. A metal-base transistor with a unipolar p-type semiconductor collector embedding InAs/GaAs quantum dots (QDs) is studied. Using minority/majority carrier injection, ballistic electron emission spectroscopy and its related hot-carrier scattering spectroscopy measures barrier heights of a buried Al$_{x}$Ga$_{1-x}$As layer in conduction band and valence band respectively, the band gap of Al$_{0.4}$Ga$_{0.6}$As is therefore determined as $2.037\pm0.009$ eV at 9 K. Under forward collector bias, interband electroluminescence is induced by the injection of minority carriers with sub-bandgap kinetic energies. Three emission peaks from InAs QDs, InAs wetting layer, and GaAs are observed in concert with minority carrier injection.

\end{abstract}


\maketitle

Utilizing hot electrons ballistically injected by a scanning tunneling microscope (STM) tip over the Schottky barrier (SB) into a semiconductor collector, ballistic electron emission microscopy (BEEM) and its associated spectroscopy (BEES) characterizes carrier filtration through buried interfaces with nanometer spatial resolution. \cite{Kaiser,PrietschVenky} BEES probes the local SB height or band offset of buried heterostructures if their depths are within the hot electron mean free path. One limitation of the BEEM technique is that it is only sensitive to transmission over energy barriers\cite{Oshea} or through quantum-confined energy levels formed between \emph{transparent} barriers.\cite{Rubin} In the active regions of typical optoelectronic devices, i.e. light-emitting diodes or lasers, quantum-confined energy levels are formed between thick barriers, which can be studied using techniques probing interband optical transitions such as electroluminescence (EL) and photoluminescence (PL). However, majority carrier injection was adopted in typical BEEM measurements, e.g. electrons into an n-type collector, making it impossible to study interband optical transitions.

Recently, the concept of ballistic electron emission luminescence (BEEL), a combination of BEEM with EL technique, has been proposed to study hot carrier transport and spontaneous emission of direct-gap semiconductor heterostructures simultaneously.\cite{Ian1,Ian2,Wei1,Wei2} In previous reports, a \emph{bipolar} light-emitting-diode-like \emph{p-i-n} heterostructure collector was used. Ballistic electrons are tunnel injected into a lightly \emph{n}-doped surface layer which ensures a SB formation with the metal base. The \emph{p-i-n} doping profile produces a built-in energy barrier at equilibrium that prevents electrons from entering the hole-rich recombination region. It was found that a threshold forward collector bias larger than the energy discrepancy between the emitted photons and the SB height was needed to induce luminescence, as required by total energy conservation if any nonlinear effects can be ignored.\cite{Kasey} In contrast to two-terminal STM-induced luminescence (STL) \cite{Gimzewski} which always requires a tip bias larger than the interband transition energies, three-terminal BEEL enables simultaneous measurement of SB heights with sub-bandgap energies.

Here we extend the scheme of BEEL on more generalized device structures. InAs QDs self-assembled in the Stranski-Krastanow growth mode on GaAs substrate is chosen as the material system because of its importance for electronic and photonic devices and relatively matured growth techniques. BEEL spectroscopy of this material system has been reported previously.\cite{Wei1} However, a bipolar collector doping profile was used, which introduces a \emph{quadratic} band bending and hence a linear electric field in the n-type epilayers above the QDs. This is generally undesirable for BEEM measurements of buried barrier heights. In this article, a \emph{unipolar} collector heterostructure is designed, with a layer of self-assembled QDs embedded in unintentionally-doped epilayers grown on p-type substrate. Such a design gives rise to a \emph{linear} band profile and hence a constant electric field in the undoped layers that can be adjusted by a collector bias (see Fig.~1). Importantly, a nonequilibrium flat-band condition can be reached accommodating BEEM measurement of buried barrier heights. Moreover, holes can be tunnel injected into the QDs from the p-doped substrate through the triangle barrier underneath, and interband optical transitions can be induced by ballistic injection of minority electrons through the SB interface into the QDs. A key part of the present design is a GaAs/AlGaAs short-period superlattice blocking layer grown above the QDs to reduce internal majority (hole) current flowing to the surface under forward collector bias, which ensures the simultaneous measurement of externally-injected minority (electron) current. The base-collector structure resembles a field-effect transistor (FET). The Schottky gate contact can be used to controllably inject single carriers into the QDs.\cite{Petroff1} Previous optical studies such as micro-PL \cite{Petroff2} and absorption spectroscopy \cite{Petroff3} have revealed rich information about the quantum-confined energy states of QDs. It is valuable to study optical transitions of these quantum states with electrical pumping to gain more insights of physical processes involved.

A single layer of InAs QDs are grown via molecular beam epitaxy on a Zn-doped \emph{p}-GaAs (100) substrate following a 15 nm thick GaAs spacer using ``partially-capped island'' (PCI) technique.\cite{Garcia,Note1} The QDs are capped by another 15 nm of GaAs followed by the growth of ten periods of GaAs/Al$_{0.4}$Ga$_{0.6}$As superlattice with well/barrier thickness of 3 nm/2.5 nm respectively. The growth is finalized with a 5 nm GaAs cap layer to prevent surface oxidation. All the epitaxial layers are unintentionally doped. Previous atomic force microscope measurements of InAs QDs grown under similar conditions found donut-shaped QDs with typical height of about 1.5 nm and ellipsoidal base dimensions about $70\times110$ nm$^{2}$.\cite{Garcia} The ground state emission energy of QDs was found to be $\simeq1.35$ eV.\cite{Garcia} The concept of such a unipolar BEEL device is demonstrated on a solid-state metal-base transistor, in which the STM-tip/vacuum/metal-base tunnel junction for BEEM is replaced by a planar Al/AlO$_{x}$/Al tunnel junction. Wavelength-resolved optical spectroscopy is facilitated with improved signal/noise ratio owing to the much larger active area (~10$^{-4}$ cm$^{2}$), albeit at the cost of spatial resolution.\cite{Wei1} Al/AlO$_{x}$/Al tunnel junctions are fabricated using a shadow-mask technique with details described previously.\cite{Ian1} Devices made from adjacent regions of the same wafer have shown similar characteristics. All the data presented is from the same device for consistency. The measurement was done at 9 K in a continuous-flow He cryostat to freeze out thermionic current and to improve the quantum efficiency of QD emissions. Photon signals are dispersed by a 0.25 m grating spectrograph and recorded by a thermoelectrically-cooled Si CCD camera.

Three-terminal collector current spectroscopy is first examined. Fig.~2 (a) and (b) are Gummel plots of collector current ($I_{C}$) as a function of emitter bias ($V_{E}$) measured at constant collector biases ($V_{C}$). At $V_{C}=0$ V, $I_{C}$ is found to be negative, i.e. holes are injected into the p-type collector even though hot electrons are injected into the base under negative $V_{E}$. This phenomenon was first observed by Bell \emph{et al.},\cite{Bell} which can be interpreted using the schematic band diagram in Fig.~1(a). Because of the \emph{repulsive} electric field in the conduction band of the collector, hot electrons incident on the SB interface can not be collected, instead they lose their energy and momentum via the dominant electron-electron (\emph{e-e}) scattering mechanism in the metal base. Secondary electrons are excited above the Fermi level and leave behind a distribution of hot holes, some of which may cross the SB interface and be collected by the \emph{attractive} electric field in the valence band of the collector. The situation drastically changes if $V_{C}$ approaches a threshold value $V_{TH}\approx$ 0.5 V, which corresponds to the flat-band condition that allows hot electrons be injected over the SB and the buried GaAs/AlGaAs superlattice into the p-type collector. For $V_{C}>V_{TH}$, $I_{C}$ is found to be positive, i.e. electron injection as a typical BEEM process illustrated in Fig.~1(b). Notably, Bell \emph{et al.} used an unbiased collector which only allows majority carrier injection irrespective of the polarity of $V_{E}$,\cite{Bell} in our case a biased collector allows either majority or minority carriers to be injected under different $V_{C}$, facilitating the occurrence of radiative recombination processes in the collector with minority carrier injection.

A quantitative analysis of the collector current spectroscopy reveals that the largest barrier heights in both conduction band ($\phi_{C}$) and valence band ($\phi_{V}$) inside the collector can be measured using minority/majority carrier injection. In our case these barriers are formed by the AlGaAs layers. For the case of indirect majority injection by carrier-carrier scattering processes, Ref.\cite{Bell} pointed out that the near-threshold yield is proportional to the power of four of the excess kinetic energy, i.e. $(eV_{E}-\phi_{V})^{4}$, rather than a quadratic function for direct minority injection by BEEM processes. For solid-state devices, the yield is the collector transfer ratio $\alpha\equiv I_{C}/I_{E}$. In Fig.~2(c), the fourth root of $\alpha$ measured at $V_{C}=0$ indeed increases linearly as $V_{E}$ goes beyond a threshold value near 0.8 V. The least squares fitting by an empirical function $\alpha=\alpha_{0}(eV_{E}-\phi_{V})^{4}$ gives $\phi_{V}=0.813\pm0.001$ eV. For the case of direct minority injection (Fig.~2(d)), the yield trace can be fitted by the prevailing two-valley Bell-Kaiser (BK) BEEM model,\cite{Oshea} which gives excellent agreement with the data.\cite{Note2} The two valleys used in the fitting are Al$_{0.4}$Ga$_{0.6}$As $\Gamma$ and GaAs X valleys with effective mass of 0.1 m$_{0}$ and 0.23 m$_{0}$ respectively,\cite{Oshea} here m$_{0}$ is the free electron mass. The threshold of AlGaAs $\Gamma$ ($\phi_{C}$) is found to be $1.224\pm0.008$ eV, which is close to the sum of the Al/GaAs SB height ($\sim$0.9 eV\cite{Yithesis}) and the GaAs/Al$_{0.4}$Ga$_{0.6}$As conduction band offset ($\sim$0.3 eV). The threshold of GaAs X is found to be $1.403\pm0.019$ eV, which is closed to the reported value.\cite{Oshea} This higher-lying valley only changes the lineshape but does not affect the turn-on threshold of $I_{C}$. The band gap of Al$_{0.4}$Ga$_{0.6}$As is therefore determined as $\phi_{C}+\phi_{V}=2.037\pm0.009$ eV, which is in excellent agreement with the accepted value of 2.040 eV at 9 K.\cite{NSM} It is noted that the resonant tunneling through the GaAs/AlGaAs superlattice miniband was not observed, since its current level of sub-pA is too small compared with the background.\cite{Smoliner} The results show that it is viable to determine energy barriers in conduction band and valence band on the same device using ballistic carrier injections and deduce the band gap of the barrier material. In principle, this can be done by conventional BEEM scheme relying only on majority injection by measuring Schottky diodes fabricated on both p-type and n-type substrates. However, it inevitably introduces experimental error due to possibly different Fermi level pinnings. Finally, it is noted the majority current injection level is much lower than minority current injection level, which is understandable since hot holes have much shorter mean free path than hot electrons in the metal base. 

Now consider interband optical transitions expected for the case of minority injection. Shown in Fig.~3(a), three BEEL emission peaks are observed under three-terminal operations with a negative emitter bias ($\left|V_{E}\right|>1.3$ V) and a positive collector bias. These peaks are assigned to GaAs bandedge emission at 1.508 eV, InAs wetting layer (WL) emission at 1.46 eV, and a broad InAs QD emission near 1.3 eV, respectively. Without applying an emitter bias, a weak two-terminal EL is observed if the collector is forward biased higher than 1.5 V, which fits the energy conservation requirement for GaAs bandedge emission. As a comparison to electrical pumping, non-resonant PL measured at the same device area shows good agreement with the BEEL data. The excitation beam is the 632.8 nm line of a HeNe laser with a power density of 2 mW/cm$^{2}$. In contrast, PL measured on the backside of the wafer shows only a GaAs emission peak at 1.502 eV, with a skewed lineshape and a small redshift of 6 meV due to the heavy Zn doping in the substrate. The QD peak energy is close to the reported ground-state emission of InAs QDs with similar structural parameters.\cite{Garcia} All the QD PL and BEEL spectra show a broad lineshape with a full width at half maximum (FWHM) of $\sim$70 meV. Considering the large amount of QDs (10$^{5}$--10$^{6}$) involved, the large FWHM is attributed to inhomogeneous broadening by QD size fluctuations. The temperature dependence of the integrated QD PL intensity shows a thermal quenching behavior. The activation energy $\sim$150 meV  estimated from Arrhenius plot (not shown) is close to the energy difference between the GaAs band gap and the QD emission, suggesting that thermal quenching of the QD emission is due to the escape of photocarriers over the GaAs barriers via thermionic emission.\cite{Fafard}

To demonstrate that the observed luminescence signals stem from ballistic injection of minority carriers through the SB interface, the collector bias dependence of the BEEL intensity is examined. Fig.~3(b) shows three traces of integrated BEEL peak intensities from InAs QDs, InAs WL, and GaAs, compared with the collector current-voltage trace. The emitter bias is fixed at --1.8 V. No light emission is detected for $V_{C}<0.4$ V, in which case holes are injected (negative $I_{C}$). For $V_{C}\geq0.5$ V, photon intensity is essentially proportional to the magnitude of $I_{C}$, supporting that the source of luminescence is indeed from external minority injection. A close-up look at the collector bias dependence reveals more interesting details, i.e. weak emissions from InAs QDs and GaAs are observed even at $V_{C}=0.4$ V, where $I_{C}$ is about zero. This can be explained by the fact that near the flat-band condition both majority and minority injections as two competitive processes occur simultaneously, nulling out the overall collector current but still cause interband radiative recombinations. Surprisingly, the InAs WL emission turns on at a higher threshold of $V_{C}=0.5$ V, beyond which minority injection of electrons dominates. The higher turn-on threshold for WL is not fully understood presently, which may be related to the carrier transfer process from quantum-well states of WL to ground states of QDs assisted by the remaining electric field. Finally, Fig.~3(c) shows the collector current dependence of integrated BEEL peak intensities. The dashed lines are linear fits of the data with a fitting range of $10^{-8}$--$10^{-6}$ A. The nearly unity slopes indicate that the luminescence intensities are essentially proportional to the number of minority carriers injected.

To illustrate the correlation between collector current and light emission, in Fig.~4 we show three-dimensional surface plots of both integrated intensity of QD emission and $I_{C}$ as a function of emitter and collector biases. For minority injection ($V_{C}\geq0.5$ V), the $V_{E}$ dependence of $I_{C}$ measured at constant $V_{C}$ has a BEES lineshape, i.e a superliner increase beyond a turn-on threshold of $V_{E}\approx-1.2$ V. For majority injection ($V_{C}<0.4$ V), $I_{C}$ becomes negative. No light emission is observed since majority carriers are injected. At constant $V_{E}$, traces of $I_{C}$ versus $V_{C}$ surge for $V_{C}\geq0.5$ V in concert with the luminescence intensity. Noticeably, no light emission is observed from the internal majority hole current, which manifests as the background of $I_{C}$ without external injection ($\left|eV_{E}\right|<\phi_{C}$) under large enough forward $V_{C}$. As a linear energy-conversion device, it is found that the total energy input ($\left|eV_{E}\right|+\left|eV_{C}\right|$) needs to be greater than 1.8 eV to generate GaAs band-edge emission at 1.51 eV. This is different than the previously reported bipolar BEEL devices, where the overall bias threshold was found to be about the same as the GaAs emission energy.\cite{Ian1} The extra energy loss of $\sim$300 meV coincides with the conduction band offset of the GaAs/Al$_{0.4}$Ga$_{0.6}$As superlattice in the present design, indicating that hot electrons injected over the AlGaAs barrier are first thermalized by multiple LO phonon emissions to the GaAs conduction band before they recombine with holes. The observed energy discrepancy between the overall bias threshold and GaAs bandedge emission can therefore be used to estimate the conduction band offset of the buried GaAs/AlGaAs heterojuction.

In summary, we have presented spectroscopic measurements of a three-terminal metal-base transistor with a FET-like p-type heterostructure collector embedding optically active InAs/GaAs QDs. By adjusting the electric field in the collector with a gating voltage bias, either energetic majority carriers or minority carriers can be ballistically injected to probe the heights of subsurface energy barriers in valence band and conduction band respectively. The band gap of the corresponding material is therefore measured unambiguously. Interband electroluminescence from the QDs active region as well as GaAs is observed in concert with injection of minority carriers. Since the unipolar collector doping profile is suitable for conventional BEEM/BEES studies, by replacing the solid-state tunnel emitter with a STM tip, our method is capable of determining subsurface energy barriers and quantum-confined energy states simultaneously at a local scale, which is currently being investigated.

One of the authors (W. Y.) thanks L. R. Ram-Mohan for commenting the manuscript. This work was supported by a DARPA HUNT contract No. 222891-01 sub-award from the University of Illinois at Urbana-Champaign, the NSF-funded Nanoscale Science and Engineering Center (NSEC), and the Center for Nanoscale Systems (CNS) at Harvard University.

\newpage

{\Large {Figure Captions}}

\vspace{20pt}

Fig.~1: (Color online) Schematic band diagrams of a metal-base transistor with the unipolar p-type collector unbiased (a), or forward biased (b). Hot electrons (blue dots) are tunnel injected into the metal base under negative $V_{E}$. For majority injection (a), secondary electrons are excited by \emph{e-e} scattering (1) and hot holes (red dots) are produced and injected over a barrier $ \phi_{V}$ into the collector (2); for minority injection (b), hot electrons are directly injected over a barrier $ \phi_{C}$ into the collector and induce interband electroluminescence.

\vspace{20pt}

Fig.~2: (Color online) Collector current spectra measured at constant $V_{C}$ for the case of majority injection (a) and minority injection (b). The corresponding collector transfer ratio $\alpha\equiv I_{C}/I_{E}$ at $V_{C}=0$ V and $V_{C}=0.65$ V are plotted in (c) and (d), respectively.\cite{Note2} The results of the two-valley BK fits are shown in (d). The numbers in the legend are the thresholds (indicated by arrows) and the attenuation factor R. The two valleys considered for the conduction band barriers are the AlGaAs $\Gamma$ (V1) and GaAs X (V2), fitted as the red dotted line and the blue dashed line respectively.\cite{Oshea} The red solid line in (c) is a least squares fitting of $\alpha$ by function $\alpha=\alpha_{0}(eV_{E}-\phi_{V})^{4}$.

\vspace{20pt}

Fig.~3: (Color online) (a)Wavelength spectra of luminescence by optical pumping (PL: blue line) and electrical pumping (BEEL: red line; two-terminal EL: gray line) measured on the same device. PL from the p-GaAs substrate (measured on the backside) is also shown (black line). All the data are normalized by the GaAs peak intensity for comparison. (b) The collector bias ($V_{C}$) dependence of integrated BEEL peak intensities for InAs WL (black squares), InAs QDs (red circles), and GaAs (green triangles) emissions; as well as collector current (blue dashed line). The luminescence data are normalized for comparison. The light gray area highlights where the collector current is negative (majority injection). (c) The collector current ($I_{C}$) dependence of integrated BEEL peak intensities. The dashed lines are linear fits of the data with a fitting range of $10^{-8}$--$10^{-6}$ A.

\vspace{20pt}

Fig.~4: (Color online) Emitter and collector bias dependence of collector current (a) and integrated BEEL peak intensity of InAs QDs (b).

\end{document}